# Modelling spatial distribution of defects and estimation of electrical degradation of silicon detectors in radiation fields at high luminosity[1]

**S. Lazanu**[a], **I. Lazanu**[b]

[a] National Institute for Materials Physics, POBox MG-7, Bucharest-Măgurele, Romania, e-mail: lazanu@infim.ro

[b] University of Bucharest, Faculty of Physics, POBox MG-11, Bucharest-Măgurele, Romania, e-mail: i_lazanu@yahoo.co.uk

**Abstract**

The irradiation represents a useful tool for determining the characteristics of defects in semiconductors as well as a method to evaluate their degradation, fact with important technological consequences.

In this contribution, starting from available data on the degradation of silicon detector characteristics in radiation fields, these effects are explained in the frame of a model that supposes also the production of the $Si_{FFCD}$ defect due to irradiation. The displacement threshold energies – different for different crystallographic axes, considered as parameters of the model, are established and the results obtained could contribute to clarify these controversial aspects.

Predictions of the degradation of electrical parameters (leakage current, effective carrier concentration and effective trapping probabilities for electrons and holes) of DOFZ silicon detectors in the hadron background of the LHC accelerator, supposing operation at -10$^0$C are done.

The non uniformity of the rate of production of primary defects and of complexes, as a function of depth, for incident particles with low kinetic energy was obtained by simulations in some particular and very simplifying assumptions, suggesting the possible important contribution of the low energy component of the background spectra to detector degradation.

**PACS**: 29.40.-n: Radiation detectors

**Key words**: detectors, silicon, modelling, damage, defects, displacement threshold energy, LHC

---

**Introduction**

Silicon detectors were widely used in experimental modern high energy physics and they will represent possible candidates for the next generation of experiments in the energy range of few to tens TeV. They are elements of the high resolution vertex and tracking system, as well as of calorimeters. The high energy radiation existent in these environments will represent a major problem for long time operation. So, this investigation could represent a very useful tool for new materials and device geometries able to work in hostile radiation environments.
In this contribution some aspects of the degradation of semiconductor material and of devices in radiation fields are investigated in the frame of a model that considers also the generation of the $Si_{FFCD}$ defect in silicon due to irradiation. Using the anisotropy of degradation correlated with the directions of crystallographic axes, values for the displacement threshold energy are established using the available data of the degradation due to irradiation on $N_{eff}$ and on the effective trapping probabilities of charge carriers for FZ and DOFZ materials. Predictions of the degradation of electrical parameters: leakage current, $N_{eff}$ and effective trapping probabilities for charge carriers for DOFZ silicon detectors in the hadron background of the LHC, at $-10^0$C are done.
A non uniformity in the rate of production of defects as a function of depth for incident particles with low kinetic energy is also predicted in some particular and very simplified cases.

**Model and results**

In a previous paper [1] the authors suggested that in the understanding of fundamental phenomena of degradation in silicon after irradiation, the consideration of the existence of the fourfold coordinated defect [2], as a primary defect, in addition to "classical" vacancies and interstitials, is essential because it has energy level(s) in the in the region of the middle of the bandgap. This defect introduces a new type of symmetry of the lattice. Fedina and co-workers [3] claimed that *in situ* HREM irradiation experiments put in evidence the existence of $Si_{FFCD}$ assigned to the known {113} defect.

Lattice degradation due to irradiation is a complicated process. A basic quantity used to describe the response of semiconductor crystalline lattice to particle irradiation is the threshold energy for atomic displacements ($E_d$). $E_d$ represents the minimum energy which leads to the formation of a lattice defect that survives at least the energy thermalization phase, when released to a target atom embedded in a crystalline host, lasting about $10^{-12}$ s [4].
The experimental determination of $E_d$ is a difficult task, since primary atomic displacements occur during very fast non-equilibrium events. The experimental detection of defects typically occurs at times much longer than $10^{-12}$ s after the collision event, when a large part of the damage may have annihilated. So, the measured $E_d$ represents an effective value.
Experimental values between 11 and 40 eV have been reported, see for example [5] and the value depends on crystal symmetry and orientation, direction of the recoil in the lattice (which is function of the direction of the irradiation beam in respect to the axes of symmetry of the crystal, energy and the mass of incident particle) and temperature.
The theoretical situation is not clearer, numerical results for $E_d$ in Si being spread in a wide range, and with values for ratio $E_d^{<111>}/E_d^{<100>} \lesssim 1$ depending on the choice of the interatomic potential and on the dimension of the cells considered in the calculations. Thus, the concrete values for $E_d$ represent a controversial problem.

In some previous investigations [6, 7] the validity and the performances of the model have been confirmed experimentally considering different situations: various damage rates, impurity concentrations and material resistivities, the contribution of different complex defects, the dependence on fluence, temperature and time, or material orientation on the degradation of leakage current, $N_{eff}$, and the effective trapping probabilities for charge carriers.

Considering the contribution of the $Si_{FFCD}$ defect to the degradation of material and device parameters in radiation field, we analyse the effects of the displacement threshold energy on $N_{eff}$, and on the effective trapping probabilities of charge carriers for FZ and DOFZ materials with <100> and <111> orientations, using available experimental data [8, 9]. Values of effective threshold energies for displacements of 19 and 26 eV give reasonable agreement with experimental data for <111> and <100> silicon irradiated samples respectively and improve the previous results of the authors.

The degradation of the electrical parameters in conditions of continuous long time irradiation – as in the future experiments at high luminosities as LHC, is modelled, considering both an infinite and a finite lifetime for $Si_{FFCD}$ defect. The radiation field is considered in agreement with the simulation presented in Reference [10], at 20 cm from the interaction point. The results are presented in Figure 1. The calculations have been performed for both threshold energies established previously, 19 and 26 eV. A major difficulty in the correct establishment of the lifetime for $Si_{FFCD}$ defect is the absence of experimental data on time dependence of the degradation of the electrical device parameters. Continuous curves correspond to infinite lifetime of the $Si_{FFCD}$ defect, while dotted ones have been calculated for a lifetime of the order $10^6$ sec at room temperature, in agreement with Ref. [11] for I-V complexes.

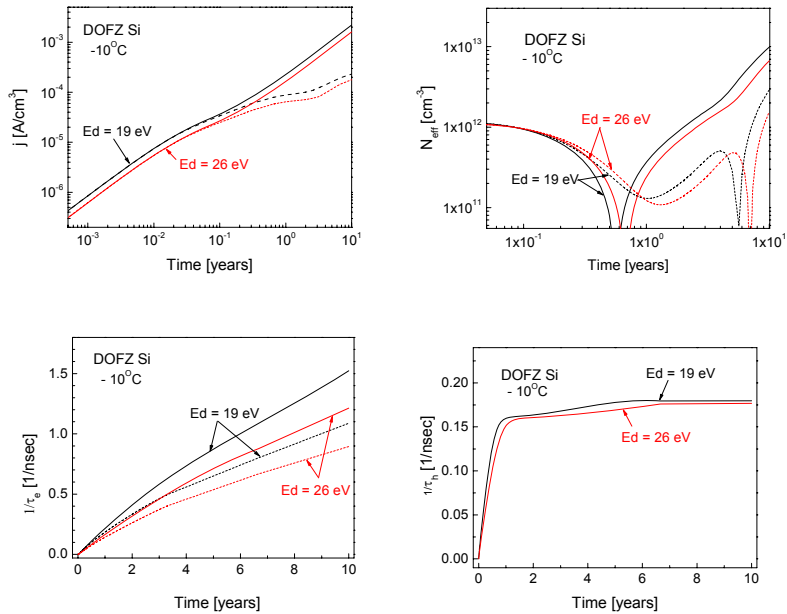

**Figure 1**
Predicted time degradation of the electrical parameters of DOFZ detectors operating in the LHC conditions at $-10^0$ C:
(a) leakage current, (b) effective carrier concentration (c and d) effective trapping probabilities of charge carriers:
continuous line-infinite lifetime of $Si_{FFCD}$; dashed line-finite lifetime (see text)

Another aspect investigated is related to the degradation of silicon due to particles with low energies. In the concrete facilities, as LHC and its up-grade as SLHC, a complex background of radiation exists in the cavity where the detection systems are placed. Currently, the low energy component of the background spectra is cut in the evaluation of the degradation. A simple analysis of CPD (concentration of primary defects) in silicon due to different incident particles for energies below 20÷30 MeV suggests that the contribution of protons is higher than the contribution due other particles in same energy region. In these conditions, although in the simulated spectra the low energy region is characterised by lower values, these protons are very damaging, and cannot be removed from any analysis.

The production of primary defects in silicon (a 300 µm thickness was considered) due to low kinetic energy (7 ÷ 50 MeV) protons was MC simulated using SRIM - 2006.02 version [12]. Preliminary results presented here are based on a simple model which supposes that the impurities (only oxygen, carbon and phosphorus are considered) are uniformly distributed in the bulk of the silicon sample and the diffusion is neglected. An analysis of displacement production in the silicon bulk as a function of the kinetic energy of the incoming protons, starting from the lowest energies for which the protons exit the detector of considered thickness (7 MeV), reveals a non uniform rate of production of defects as a function of depth for incident protons with energies up to 15 MeV; at higher energies practically this phenomenon disappears. This effect is also influenced by the anisotropy of the threshold energy for displacements for different directions, but this aspect has not been studied here. Also, the existence of a difference between surface and bulk values for the threshold energy for atomic displacement must be mentioned. The rule ½ is the simplest approximation but this problem remains open. The following predefined SRIM values have been used: displacement energy -15eV, surface binding energy - 4.7eV, lattice binding energy – 2 eV. The distribution of vacancies and interstitials obtained by MC simulation has been used to obtain the distribution of complex defects: $V_2$, VO, VP, $C_i$, $C_iO_i$ and $C_iC_s$. The results obtained for 7 MeV incident protons in these simplifying hypotheses show (see Figure 2) that at low fluences, up to $10^{11}$ p/cm$^2$, defect distribution follows the distribution of primary defects, and the relative contribution of each complex depends on the initial concentration of impurities and on the energy of formation of defects. The increase of the particle fluence conduces to a saturation of the volume concentration of defects related to impurities, limited by the initial concentrations of impurities.

Further work with more realistic hypotheses is necessary.

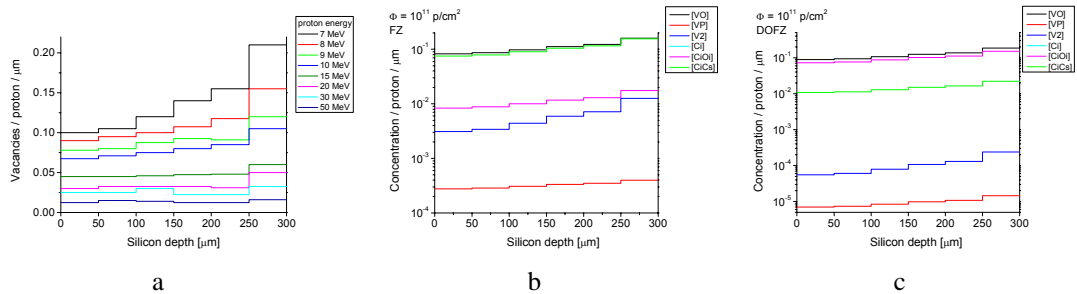

Figure 2
The simulated production of primary defects in silicon of 300 μm thickness due to protons with 7 MeV kinetic energy (a) and the distribution of complex defects in FZ material (b) and DOFZ (c) respectively.

## Summary

The consideration of a finite lifetime for the $Si_{FFCD}$ defect makes the model more realistic, but measurements of the detector characteristics over long time intervals, at room temperature, are necessary for the clarification of this aspect.

Values of effective threshold energies for displacements of 19 and 26 eV give a reasonable agreement with experimental data for <111> and <100> silicon irradiated samples respectively.

In the hypothesis considered in the simulation of the distribution of primary defects produced by low energy protons in the depth of silicon samples, a non-uniformity of damage in the depth has been obtained and this is maximised for particles which have a range equal to the thickness of the sample. The maximum of the damage is located at the side where the particle gets out. The increase of the energy of the incoming particles produces an increase in uniformity of the damage, and a decrease of its magnitude.

## Acknowledgments

One of the authors (S.L.) wishes to thank the organisers for the warm hospitality and for financial support.